\documentclass[a4paper]{article}
\pdfoutput=1
\usepackage{amsfonts}
\usepackage{amssymb}
\usepackage{amsmath}
\usepackage{subfigure}
\usepackage{graphicx}
\usepackage{authblk}
\usepackage{color}

\begin{document}

\title{Random time-inhomogeneous Markov chains}
\author[1]{G.C.P. Innocentini \thanks{ginnocentini@gmail.com}}
\author[2]{M. Novaes \thanks{mnovaes@ufu.br}}

\affil[1]{Federal University of ABC, Brazil}
\affil[1,2]{Instituto de F{\'i}sica,  Universidade Federal de Uberl{\^a}ndia, Brazil}

\date{\today}

\maketitle
\begin{abstract}
We consider Markov chains with random transition probabilities which, moreover, fluctuate randomly with time. We describe such a system by a product of stochastic matrices, $U(t)=M_t\cdots M_1$, with the factors $M_i$ drawn independently from an ensemble of random Markov matrices, whose columns are independent Dirichlet random variables. The statistical properties of the columns of $U(t)$, its largest eigenvalue and its spectrum are obtained exactly for $N=2$ and numerically investigated for general $N$. For large $t$, the columns are Dirichlet-distributed, however the distribution is different from the initial one. As for the spectrum, we find that the eigenvalues converge to zero exponentially fast and investigate the statistics of the largest stability and Lyapunov exponents, which is are approximated by Gamma distributions. We also observe a concentration of the spectrum on the real line for large $t$.
\end{abstract}

{\bf keywords:} Markov processes, random matrices, Lyapunov exponents, products of matrices

\section{Introduction}
\indent

Markov chains are a fundamental statistical model with numerous applications, that range from computation and physics to chemistry and biology \cite{book}. We consider systems evolving in discrete time and having access to a finite number of possible different states. Let us say that there are $N$ such states and the system may be characterized by a vector $\vec{p}(t)=(p_1(t),...,p_N(t))^T$ such that $p_i(t)$ is the probability that site $i$ is populated at time $t$ (this may be e.g. the probability that a spin system has $i$ spins pointing up, or that there are $i$ individuals in a population, or that a random walker has reached a position $i$ steps away from where he started).

The system is called a Markov chain if $\vec{p}(t+1)=M_t\vec{p}(t)$ for some matrix $M_t$, the transition matrix at time $t$. The element $(M_t)_{i,j}$ is interpreted as the transition probability from state $j$ to state $i$. If we define $|\vec{p}|:=\sum_{i=1}^N p_i$, then the normalization condition $|\vec{p}(t)|=1$ must hold at all times and, therefore, the transition matrix must satisfy $\sum_{i=1}^N{(M_t)_{i,j}}=1$ for every $j$, i.e its columns must be normalized. Such matrices are called stochastic.

If the transition probabilities are constant in time, i.e. if $M_t=M_1=M$ for all $t$, the system is said to be time-homogeneous and the population probabilities evolve according to the powers of the matrix, $\vec{p}(t)=M^t\vec{p}(0)$. If $M$ is irreducible and aperiodic, the Perron-Frobenius theorem guarantees that it has a unit eigenvalue, $\lambda_0=1$, and that all other eigenvalues are strictly smaller in modulus. The unit eigenvalue has a left eigenvector given by $(1,...,1)$,  and a right eigenvector with non-negative elements, called the stationary state. Time evolution in the homogeneous case tends to the stationary state exponentially fast for almost all initial conditions, with a time scale for long times controlled by the second-largest eigenvalue, $\lambda_1$. 

An ensemble of random stochastic matrices has been introduced \cite{horvat,chafai} such that the columns are independent random variables, i.e. its elements are written as \begin{equation} M=\left( \vec{v}_1\,\cdots\,\vec{v}_N\right)\end{equation} where each vector $\vec{v}_j=(M_{1,j},...,M_{N,j})^T$ is taken at random with a Dirichlet distribution: \begin{equation}P_{\vec{a}}(\vec{v})=\frac{\delta(|\vec{v}|-1)}{\Gamma(a_0)}\prod_{i=1}^N \Gamma(a_i)v_i^{a_i-1},\end{equation} where $a_i>0$ and $a_0=|\vec{a}|$. The marginal distribution of the $i$th component is a Beta distribution,
\begin{equation} \frac{\Gamma(a_0)}{\Gamma(a_i)\Gamma(a_0-a_i)}v_i^{a_i-1}(1-v_i)^{a_0-a_i-1}.\end{equation}
In line with previous works, we shall consider only the simplest case in which all parameters are equal, $a_i=a$, and $a_0=Na$. This can be produced by drawing $N$ independent variables with Gamma distribution  $\frac{x^{a-1}}{\Gamma(a)}e^{-x},$ and then normalizing them. In this case all elements have the same marginal distribution, namely 
\begin{equation} \label{beta}B_a(v)=\frac{\Gamma(Na)}{\Gamma(a)\Gamma((N-1)a)}v^{a-1}(1-v)^{(N-1)a-1}.\end{equation}

Almost all such random stochastic matrices are indeed irreducible and aperiodic and hence subject to the Perron-Frobenius theorem. For large dimensions, $N\gg 1$, it is known \cite{bordenave} that the eigenvalues smaller than 1 are distributed uniformly inside a disk in the complex plane. The radius of this disk is given by the modulus of the second-largest eigenvalue, $|\lambda_1|$, which concentrates at $1/\sqrt{N}$ and seems to have \cite{horvat} a Gumbel distribution around it with variance decreading with $N$. It has also been observed that, for large $N$, each element of the Perron-Frobenius right eigenvector (associated with the unit eigenvalue) has a Gaussian distribution around the mean $1/N$, with variance decreasing as $1/N^3$ \cite{horvat}.

If the transition probabilities are constant in time, the evolution of the system is governed by the powers $M^t$, which have the same eigenvectors as $M$ andwhose eigenvalues are the eigenvalues of $M$ raised to the power $t$. The motivation of the present work is to generalize the notion of random Markov chains discussed above to include time-inhomogeneity, i.e. to consider situations in which the transition probabilities may depend on time in a way which is itself random. In this case, the evolution of population probabilities is given by 
\begin{equation}\label{mc1}
    \vec{p}(t)=M_t\cdots M_2M_1\vec{p}(0)=U(t)\vec{p}(0),
\end{equation}
with the factors $M_i$ being independent.

Let us notice that there are therefore two levels of probability at work. First, we have a random system that has probability $p_i(t)$ of being in state $i$ at time $t$. Second, the probabilities $p_i(t)$ evolve with time according to a transition matrix which is also a random variable. This might correspond, for example, to a random walker who is moving around amid a strong irregular wind, so that the probability of moving in a certain direction fluctuates randomly with time (a random walker in a random environment). Similar ideas, but in more general and more mathematically-oriented contexts, were explored for example in \cite{rcre1,rcre2,rcre3}.

We must thus consider products of random matrices. Historically, investigations about products of random matrices have mainly proceeded along two lines: the regime of large $t$ for fixed $N$, pioneered by Furstenberg and Kesten \cite{furs}, or the regime of large $N$ for fixed $t$. In the first case it is common to consider a finite sample space, and the figure of merit is either the largest stability exponent $\theta_1$, defined as $|\lambda_1(t)|\sim e^{-\theta_1 t}$ or the largest Lyapunov exponent $\vartheta_1$, defined as $z_1(t)\sim e^{-\vartheta_1 t}$ where $z_1$ is the largest singular value of $M$. It is expected that these exponents have Gaussian distributions, and this can be proved for certain classes of matrices. A review of this area has been presented in \cite{viana}. 

The second regime, sometimes motivated by the theory of free probability, tends to focus on random matrices with Gaussian elements, hermitian or not, and consider, besides those exponents, macroscopic spectral properties, although some microscopic results are also available. For reviews, see \cite{burda3,akeipsen2}. Most attention has been given to the classical ensembles of Ginibre \cite{burda1, burda2, akemann1, akemann2, ipsen1, kargin, kuijlaars,akemannf} and Wishart \cite{akemann3, akemann4}, and also to ensembles of truncated unitary matrices \cite{akemann5,kieburg}, among others \cite{soshnikov1, soshnikov2}. The investigation of products of random matrices appears in the analysis of stochastic dynamical systems where the knowledge of the distribution of stability/Lyapunov exponents is necessary to study questions of stability \cite{newman1, newman2, forrester1, forrester2}.

The problem of multiplying random stochastic matrices has some peculiarities. Our ensemble is a semi-group: the product of two stochastic matrices is also stochastic. The Perron-Frobenius theorem therefore also applies to $U(t)$, which means its largest eigenvalue is always $1$, in contrast to Gaussian random matrices, for example. The remaining eigenvalues decay exponentially with $t$ and we consider the distribution of the largest stability and Lyapunov exponents: for small $N$, they are very well described by a Gamma distribution for any $t$; on the other hand, for large $t$ they tend to have a Gaussian distribution for any $N$.

We also consider the distribution of the columns of $U(t)$. Since all but one of the eigenvalues of $U(t)$ vanish as $t\to\infty$, its columns converge towards a common limit, which coincide with the Perron-Frobenius eigenvector. For $N=2$ we prove that this limit has a Dirichlet distribution. Numerical results suggest that this remains valid for arbitrary $N$. For large $N$, the elements tend to be uncorrelated Gaussian random variables.

Finally, we look at the spectrum as a whole. We find that there is a finite fraction of real eigenvalues, as happens for real Ginibre matrices \cite{edelman}. As $t$ grows, the eigenvalues increasingly concentrate on the real line. This concentration phenomenon has also been observed for products of other kinds of random matrices \cite{forrester3,laksh,hameed,ipsen2,forr,simm,truncortho}. 

This paper is organized as follows. We first study the simplest problem of $2\times 2$ matrices in Section 2. In this context we are able to rigorously derive the distribution of the eigenvalue and of the matrix elements of $U(t)$ for large $t$. We then proceed to numerical experiments. In Section 3 we conjecture that the matrix elements of $U(t)$ satisfy, for large $t$, a Dirichlet distribution with $a'=aN$. Section 4 is dedicated to the distribution of thestability and Lyapunov exponents of $U(t)$. With all eigenvalues of $U(t)$ tending to zero as $t\to\infty$, the columns of the matrix must become equal. We measure this convergence by the difference between the first elements in each column, $d_{ij}=|U_{1i}-U_{1j}|$. This quantity decays exponentially with $t$; its distribution is independent of $i$ and $j$ and well approximated by a Gamma distribution.

\section{The case $N=2$}
\indent

In this section we restrict our attention to $N=2$. The matrices $M_i$ have independent random columns, and we start with the simplest Dirichlet distribution with parameter $a=1$. In this case, each such matrix is of the form
\begin{equation}\label{seed}
M=\left(\begin{array}{cc}
    r & s \\
    1-r & 1-s
\end{array}\right),
\end{equation}
with $r$ and $s$ uniformly distributed in the interval $[0,1]$.  

Consider the product of two such matrices,
\begin{align}
    U&=\left(\begin{array}{cc}
    r_2 & s_2 \\
    1-r_2 & 1-s_2
\end{array}\right)
\left(\begin{array}{cc}
     r_1 & s_1 \\
     1-r_1 & 1-s_1
\end{array}\right)\\\nonumber
&=\left(\begin{array}{cc}
    r_1(r_2-s_2)+s_2 &s_1(r_2-s_2)-s_2 \\
    1-r_1(r_2-s_2)-s_2 & 1-s_1(r_2-s_2)-s_2
\end{array}\right).
\end{align}
The distribution of the top left element $r_1(r_2-s_2)+s_2$, for instance, can be obtained as
\begin{equation}\label{trans2}
   P_2(U_{11}=z)=\int_{0}^{1}\int_{0}^{1}\int_{0}^{1}\delta[z-r_1(r_2-s_2)-s_2]dr_1dr_2ds_2,
\end{equation}
which gives 
\begin{equation}\label{trans4}
    P_2(U_{11}=z)=-2z\log(z)-2(1-z)\log(1-z).
\end{equation}

\begin{figure}[tb]
\begin{center}
\subfigure[]{
\includegraphics[width=2.5in,angle=0]{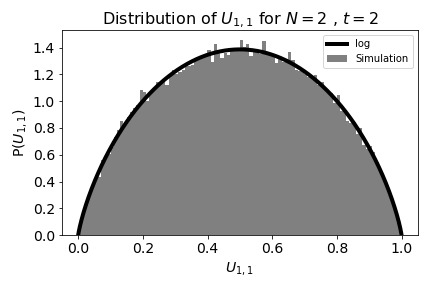}\label{fig1a}}
\subfigure[]{
\includegraphics[width=2.5in,angle=0]{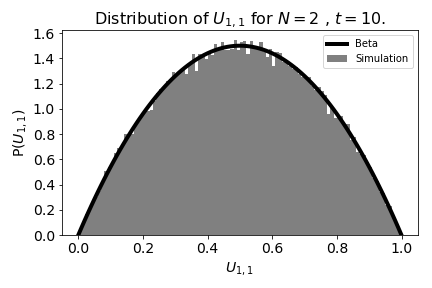}\label{fig1b}}\\
\subfigure[]{
\includegraphics[width=2.5in,angle=0]{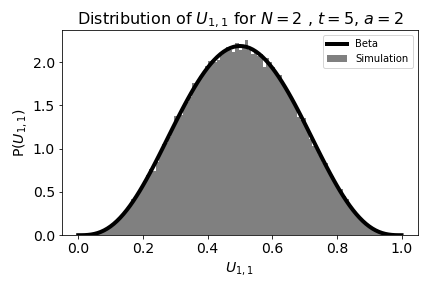}\label{fig1c}}
\subfigure[]{
\includegraphics[width=2.5in,angle=0]{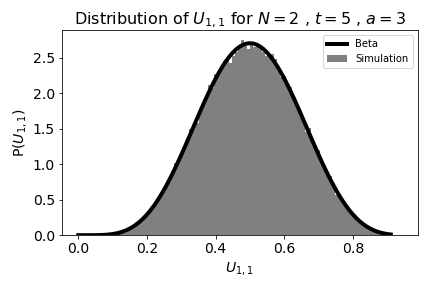}\label{fig1d}}
\caption{\label{fig1} Comparison between analytical result and numerical simulations. Fig.\ref{fig1a} and Fig.\ref{fig1b} show the probability distribution for the element $U_{1,1}$ at instants of time $t=2$ and $t=10$, respectively, when the columns have a Dirichlet density with parameter $a=1$. Fig.\ref{fig1c} and Fig.\ref{fig1d} show the probability distribution for the element $U_{1,1}$ at $t=5$ when the columns have a Dirichlet density with parameter $a=2$ and $a=3$, respectively.}
\end{center}
\end{figure}

In general, let $P_t$ denote the distribution of the top left element of $U(t)$. Then, we have the relation
\begin{equation}\label{asym1}
   P_t(z)=\int_{0}^{1}\int_{0}^{1}\int_{0}^{1}P_{t-1}(w)\delta[z-w(r-s)-s]dwdrds,
\end{equation} which can be iterated for any $t$ to produce exact yet cumbersome results. Let us instead inquire about its asymptotic behavior for large $t$. If $P_t$ is to converge to some well defined function $P_\infty(z)$, then it must be a fixed point of the iteration, i.e. it must satisfy the integral equation 
\begin{equation}
   P_\infty(z)=\int_{0}^{1}\int_{0}^{1}\int_{0}^{1}P_{\infty}(w)\delta[z-w(r-s)-s]dwdrds.
\end{equation}

It is easy to see that the function $P_\infty(z)=6z(1-z)$ satisfies this equation. By symmetry, every element of $U$ must be distributed according to this density for large $t$. This is a particular case of the Beta distribution $B_{a'}$ given in (\ref{beta}), corresponding to the parameter $a'=2$. This in turn implies a Dirichlet distribution for the columns of $U(\infty)$ with parameter $a'=2$. In Figure \ref{fig1}, we compare this analytical result with numerical simulations for $t=10$. We see that even for small values of $t$ the asymptotic result agrees very well with the data.

This approach can be applied to a more general Dirichlet distribution for the factors, with $a\neq 1$. In this case, we consider the following integral
\begin{equation}
   P_{t}(z)=\int_{0}^{1}\int_{0}^{1}\int_{0}^{1}P_{t-1}(w)P_1(r)P_1(s)\delta[z-w(r-s)-s]dwdrds,
\end{equation}
where the distributions $P_1(r)$ and $P_1(s)$ are no longer constant but given by (\ref{beta}). The fixed-point condition is
\begin{equation}
   P_{\infty}(z)=\int_{0}^{1}\int_{0}^{1}\int_{0}^{1}P_{\infty}(w)P_1(r)P_1(s)\delta[z-w(r-s)-s]dwdrds
\end{equation}
and we prove in the appendix that the function $P_\infty(z)=\frac{\Gamma(4a)}{\Gamma(2a)^2}z^{2a-1}(1-z)^{2a-1}$  satisfies it. We also show in Figure \ref{fig1} the comparison between simulation and this result for $a=2$ and $a=3$. 

\begin{figure}[tb]
\begin{center}
\subfigure[]{
\includegraphics[width=2.5in,angle=0]{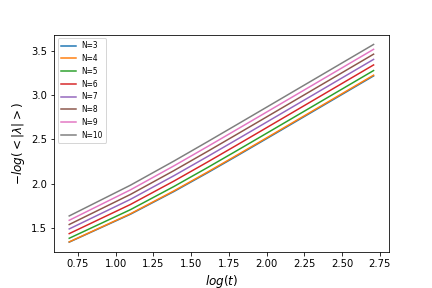}\label{fig2a}}
\subfigure[]{
\includegraphics[width=2.5in,angle=0]{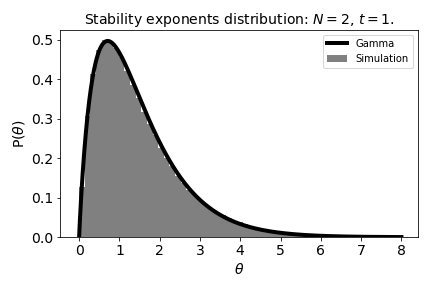}\label{fig2b}}\\
\subfigure[]{
\includegraphics[width=2.5in,angle=0]{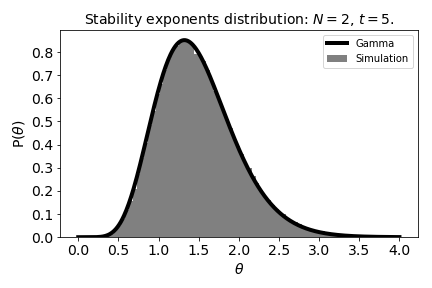}\label{fig2c}}
\subfigure[]{
\includegraphics[width=2.5in,angle=0]{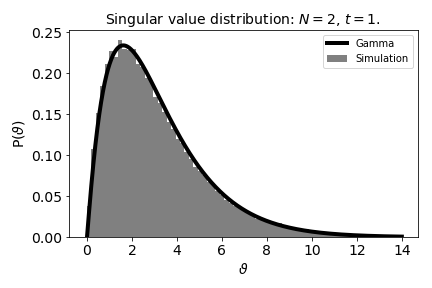}\label{fig2d}}
\caption{\label{fig2} In Fig. \ref{fig2a} we show the behavior of the $-\log(\langle \left| \lambda(t) \right| \rangle)$ as a function of time for different dimensions of the matrix $U(t)$, as indicated in the figure. In panels b) and c) we exhibit the comparison between simulation results for $N=2$ and a fitted Gamma distribution, Eq. (\ref{gamma2}), for the random variable $\theta=-\frac{1}{t}\log(\langle|\lambda_1|\rangle)$ (the parameters of the Gamma distribution are $\alpha=1.92,\beta=1.3)$ for $t=1$ and $\alpha=9.13,\beta=6.15$ for $t=5$). In panel d) we compare a fitted Gamma distribution (parameters $\alpha=2.05,\beta=0.65$) with the numerical result for the largest Lyapunov exponent $\vartheta$.}
\end{center}
\end{figure}

The largest eigenvalue of $U(t)$ is always 1, and we expect the modulus of the other one, $|\lambda(t)|$, to decay exponentially fast. This is confirmed in panel a) of Figure \ref{fig2}, for several values of $N$. We therefore consider the distribution of the stability exponent, $\theta=-\frac{1}{t}\log\langle \left| \lambda(t) \right| \rangle$. In panels b) and c) of Figure \ref{fig2} we show that this is very well described by a Gamma distribution \begin{equation}\label{gamma2} \frac{\beta^\alpha}{\Gamma(\alpha)}x^{\alpha-1}e^{-\beta x},\end{equation} for all values of $t$. The same is true for the Lyapunov exponent $\vartheta=-\frac{1}{t}\log\langle z(t) \rangle$, where $z(t)$ is the largest singular value of $U(t)$, which is shown in panel d). Both distributions approach a Gaussian for large $t$.

\section{Arbitrary $N$}
\indent

\subsection{Columns}
\indent

\begin{figure}[!htb]
\begin{center}
\subfigure[]{
\includegraphics[width=2.5in,angle=0]{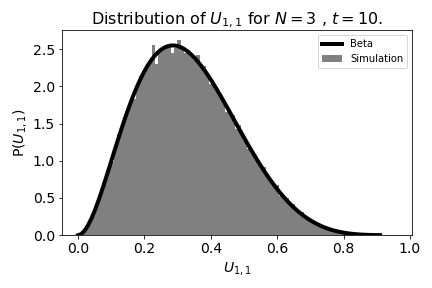}\label{fig3a}}
\subfigure[]{
\includegraphics[width=2.5in,angle=0]{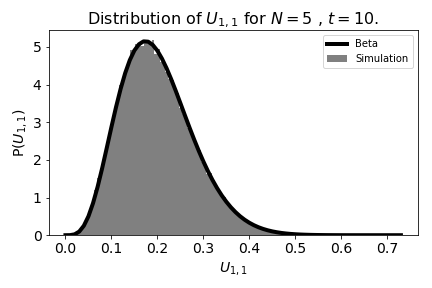}\label{fig3b}}\\
\subfigure[]{
\includegraphics[width=2.5in,angle=0]{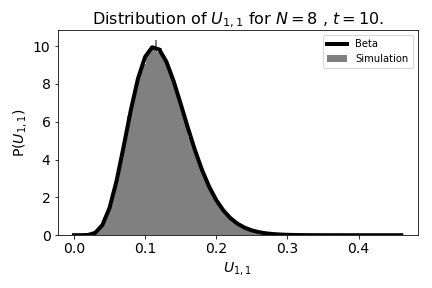}\label{fig3c}}
\subfigure[]{
\includegraphics[width=2.5in,angle=0]{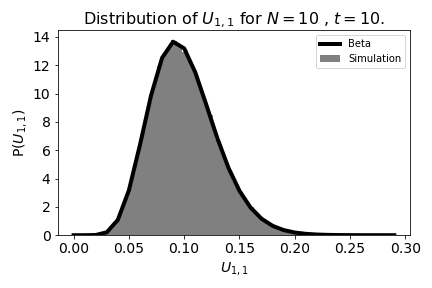}\label{fig3d}}\\
\subfigure[]{
\includegraphics[width=2.5in,angle=0]{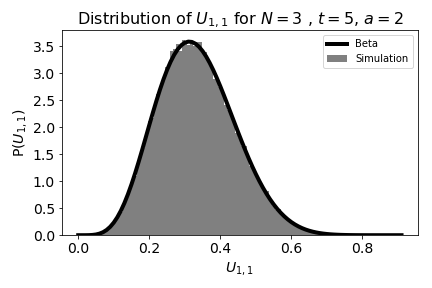}\label{fig3e}}
\subfigure[]{
\includegraphics[width=2.5in,angle=0]{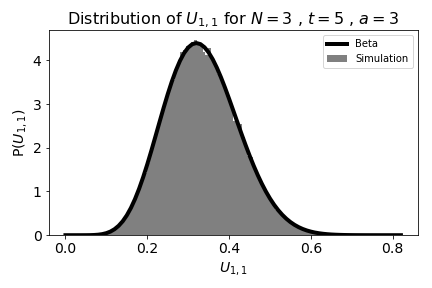}\label{fig3f}}
\caption{\label{fig3} a)-d): Distribution of $U_{1,1}(t)$ for different values of $N$ and $t$, when the columns of the factors $M_i$ have a Dirichlet density with parameter $a=1$ (panels a-d), $a=2$ (panel e) and $a=3$ (panel f). The solid line is our conjecture, Eq. (\ref{Na}).}
\end{center}
\end{figure}

\begin{figure}[!htb]
\begin{center}
\subfigure[]{
\includegraphics[width=2.5in,angle=0]{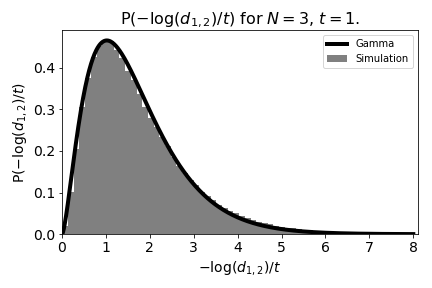}\label{fig4a}}
\subfigure[]{
\includegraphics[width=2.5in,angle=0]{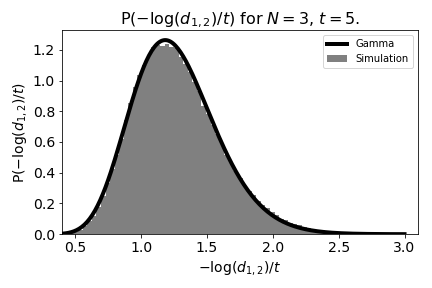}\label{fig4b}}\\
\end{center}
\begin{center}
\subfigure[]{
\includegraphics[width=2.5in,angle=0]{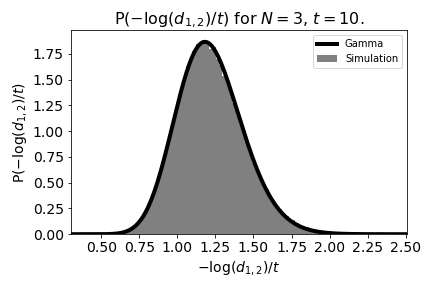}\label{fig4c}}
\caption{\label{fig4} Distribution of $P(-\frac{1}{t}\log(d_{1,2}))$, where $d_{i,j}=|U_{1,i}-U_{1,j}|$ measures the distance between columns of $U$, for $N=3$.}
\end{center}
\end{figure}

We now consider the distribution of columns of $U(t)$ for more general dimensions. We saw in the previous section that, for large $t$ and $N=2$, they satisfy a Dirichlet distribution with parameter $a'=2a$. We conjecture that, for arbitrary $N$, they still satisfy a Dirichlet distribution for large $t$, with a new parameter which is $a'=Na$. This implies a marginal distribution for the elements which is \begin{equation}\label{Na} \frac{\Gamma(N^2a)}{\Gamma(Na)\Gamma(N(N-1)a)}z^{Na-1}(1-z)^{N(N-1)a-1}.\end{equation} We cannot prove this result for $N>2$, but present in Figure 3 some numerical results for different values of $N$ which support it.

Since all eigenvalues but one vanish as $t\to\infty$, the columns of $U$ must approach a common limit. A matrix whose columns are all identical will have that very column as eigenvector. Therefore, the Perron-Frobenius eigenvector of $U(t)$ also has a Dirichlet distribution with parameter $Na$.

For large $N$, such a Dirichlet distribution converges to a Gaussian with average value of $1/N$, which of course is to be expected, and variance given by $1/(aN^3)$. Previous works have found that for $t=1$ and $a=1$ the Perron-Frobenius eigenvector was Gaussian distributed with average $1/N$ and variance $\sim N^{-3}$ \cite{horvat}. Our results suggest this behaviour extends to $U(t)$ for large $t$ and arbitrary $a$.

The rate of approach of the columns of $U(t)$ towards a common limit can be measured by considering the distribution of $d_{i,j}=|U_{1,i}-U_{1,j}|$. We have checked that this decays exponentially, and in Figure \ref{fig4} we show the distribution $P(-\frac{1}{t}\log(d_{1,2}))$ (it is independent of the particular choice of $i$ and $j$). It is very well approximated by a Gamma distribution, Eq. (\ref{gamma2}), with parameters which are non-trivial functions of $N$ and $t$. For large $t$, the distribution becomes Gaussian.

\subsection{Exponents}
\indent

\begin{figure}[!htb]
\begin{center}
\subfigure[]{
\includegraphics[width=2.5in,angle=0]{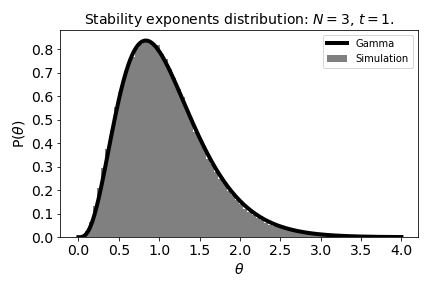}\label{fig5a}}
\subfigure[]{
\includegraphics[width=2.5in,angle=0]{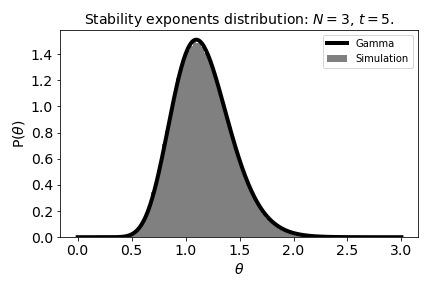}\label{fig5b}}\\
\subfigure[]{
\includegraphics[width=2.5in,angle=0]{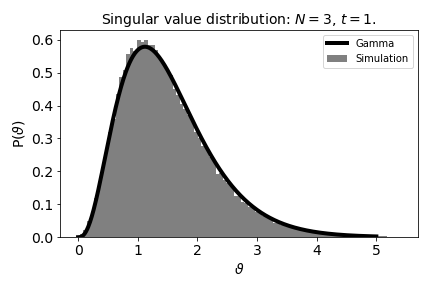}\label{fig5c}}
\subfigure[]{
\includegraphics[width=2.5in,angle=0]{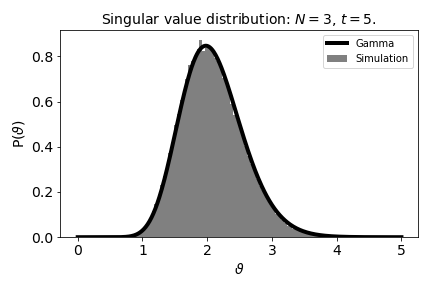}\label{fig5d}}
\caption{\label{fig5} Distribution of the largest stability exponent $\theta$, in panels a) and b), compared to a fitted Gamma distribution, Eq.(\ref{gamma2}) (parameters $\alpha=4.23,\beta=3.87$ for $t=1$ and $\alpha=18.46,\beta=15.89$ for $t=5$). In panel c) and d), distribution of the largest Lyapunov exponent $\vartheta$, also compared to a fitted Gamma distribution (parameters $ \alpha=3.8,\beta=2.5$ for $t=1$ and $\alpha=18.81,\beta=9$ for $t=5$).}
\end{center}
\end{figure}

For general $N$, the largest stability exponent is again defined as $\theta=-\frac{1}{t}\log(|\lambda_1(t)|)$, where $\lambda_1$ is the second-largest eigenvalue of $U(t)$ in modulus. As in the $N=2$ case, this quantity has an approximate Gamma distribution, Eq.(\ref{gamma2}), as can be seen in Figure \ref{fig5}.a and Figure 5.b, and the parameters $\alpha(N,t)$ and $\beta(N,t)$ have a non trivial dependence on $N$ and $t$.

For large values of $t$ with fixed $N$, the distribution converges towards a Gaussian, with mean value independent of $t$ but increasing with $N$. The variance, on the other hand, decreases with both $N$ and $t$.

The largest Lyapunov exponent is $\vartheta=-\frac{1}{t}\log(z_1(t))$, where $z_1$ is the second-largest singular value of $U(t)$. This quantity also has an approximate Gamma distribution, Eq.(\ref{gamma2}), as can be seen in Figure \ref{fig5}.c and Figure 5.d. (the parameters $\alpha(N,t)$ and $\beta(N,t)$ again have a non trivial dependence on $N$ and $t$).

\subsection{Spectrum}
\indent

\begin{figure}[!htb]
\begin{center}
\subfigure[]{
\includegraphics[width=2.5in,angle=0]{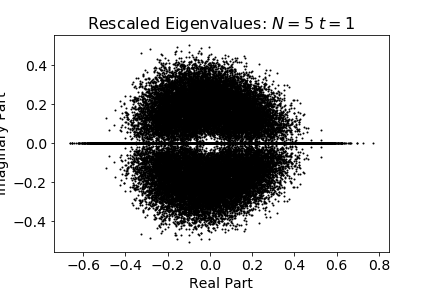}\label{fig6a}}
\subfigure[]{
\includegraphics[width=2.5in,angle=0]{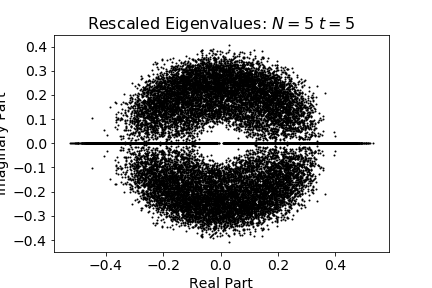}\label{fig6b}}\\
\subfigure[]{
\includegraphics[width=2.5in,angle=0]{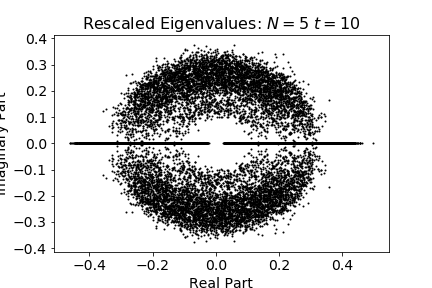}\label{fig6c}}
\subfigure[]{
\includegraphics[width=2.5in,angle=0]{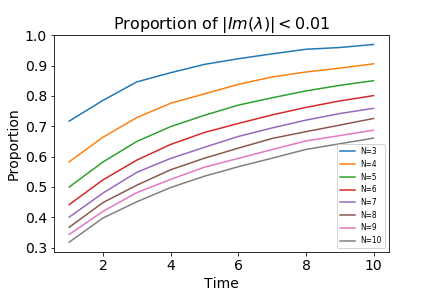}\label{fig6d}}
\caption{\label{fig6} Spectrum of the matrix $U(t)$ for $N=5$ and $t=1, 5, 10$. The bottom right panel shows the evolution of the proportion of eigenvalues with $|Im(\lambda)|<0.01$ for different values of $N$.}
\end{center}
\end{figure}

It is known that, for large $N$, real Ginibre matrices have a finite fraction of real eigenvalues (asymptotically $\sqrt{2N/\pi}$ of them \cite{edelman}). Moreover, a product of $t$ such matrices has $\sqrt{2tN/\pi}$ real eigenvalues \cite{simm}, to leading order for large $N$ and small $t$ (this $\sqrt{t}$ growth is also true for a product of truncated orthogonal matrices \cite{truncortho}). Although we are more interested in small matrices, we have observed that something similar also happens in the context we are addressing. In panel a) of Figure 6, for example, we see that for $N=5$ and $t=1$ the spectrum consists of a set of points on the real line and a cloud of complex points.

In panels b) and c) of Figure 6, we show the spectrum of $U(5)$ and $U(10)$. As the modulus of $\lambda$ tends to zero exponentially fast when $t\rightarrow\infty$, we have rescaled the eigenvalues of $U(t)$ as follows: $\lambda\to|\lambda|^{1/t-1}\lambda$. This creates a hole in the vicinity of the origin, which is an artefact. The important features are the more circular profile of the cloud of complex points, its apparent repulsion by the real line and that, as $t$ grows, the spectrum accumulates on the real line.

This concentration is quantified in Figure 6.d, where we plot the fraction of points having imaginary part in the small interval $(-0.01,0.01)$ as a function of $t$ for several values of $N$. This fraction is always increasing and seems to approach 100\% for every $N$. A direct comparison with the asymptotic Ginibre $\sqrt{2tN/\pi}$ result is not possible since we have $N=5$, but the growth in $t$ is sublinear  for small $t$, possibly as $t^{1/2}$.

\section{Conclusion}
\indent

Stochastic matrices are a very important class of matrices in physics, describing a wealth of systems with markovian dynamics. Since random matrix theory has been successfully applied in many areas of physics, it is natural to define random stochastic matrices. 

Time-homogeneous Markov chains with transition matrix $M$ are described simply by the powers $M^t$.
More general systems, however, with random time-dependent transition probabilities, evolve according to a product of random matrices, $U(t)$. This may have rather different statistics from $M^t$. 

We have analyzed in this work various statistical properties of $U(t)$. Its columns have a Beta distribution for large $t$; its second-largest stability exponent and Lyapunov exponent both have distributions that well approximated by Gamma distributions for finite $N$ and that become Gaussian for large $N$. 

Since $U(t)$ is a real non-symmetric matrix, it is understandable that it should have some similarities with real Ginibre matrices, which indeed show up in the overall distribution of eigenvalues: a finite fraction of them are real for finite $N$, and this fraction grows with $t$.

\section*{Acknowledgment}
G.C.P. Innocentini acknowledges financial support from PNPD-CAPES. M. Novaes acknowledges financial support from CNPq (grants 303634/2015-4 and 400906/2016-3) and FAPEMIG (PPM-00126-17).

\section*{Appendix}
\indent

We wish to show that the function $P_{\infty}(z)=\frac{\Gamma(4a)}{\Gamma(2a)^2}[z(1-z)]^{2a-1}$ satisfies the integral equation
\begin{equation}
   P_{\infty}(z)=\int_{0}^{1}\int_{0}^{1}\int_{0}^{1} P_{\infty}(w)P_1(r)P_1(s)\delta[z-w(r-s)-s]dwdrds,
\end{equation}
where  and $P_1(r)$ and $P_1(s)$ are the Beta distribution (\ref{beta}). We first consider a change of variables $\eta=z-w(r-s)-s$, which leads to $|dw|=|d\eta|/|r-s|$. The integral becomes:
\begin{equation}
\frac{\Gamma(4a)}{\Gamma(a)^4}\int_{0}^{1}\int_{0}^{1}\int_{\eta(0)}^{\eta(1)}\frac{[(z-\eta-s)(\eta+r-z)]^{2a-1}[r(1-r)s(1-s)]^{a-1}}{(r-s)^{4a-1}}\delta(\eta)d\eta drds,
\end{equation}
where $\eta(0)=z-s$ and $\eta(1)=z-r$ are the new limits of integration, corresponding to $w=0$ and $w=1$, respectively.  

Some caution is needed to guarantee that the integration over $\eta$ indeed contains the zero of the Dirac delta function in its range. We must impose some restriction to the values that $r$ and $s$ can assume. There are two possibilities: either 
\begin{equation}\nonumber
i)\,\, z \leq s \leq 1 \hspace{0.2cm} {\rm and} \hspace{0.2cm} 0 \leq r \leq z,
\end{equation}
or
\begin{equation}\nonumber
ii)\,\, 0 \leq s \leq z \hspace{0.2cm} {\rm and} \hspace{0.2cm} z \leq r \leq 1.
\end{equation}

In case $i)$, performing the integral over $\eta$, changing $s=r-x$ and evaluating the integrals over $x$ and $r$ we have 
\begin{equation}
  \frac{\Gamma(4a)}{\Gamma(a)^4}\int_{0}^{z}\int_{r-z}^{r-1}\frac{[(r-z-x)(r-z)]^{2a-1}[r(1-r)(r-x)(1-r+x)]^{a-1}}{x^{4a-1}}dxdr=\frac{P_{\infty}(z)}{2}.\end{equation}
   
The calculation for case $ii)$ leads to the same result as above. Therefore, when cases $i)$ and $ii)$ are combined we arrive at the desired result.

\end{document}